\documentclass[ reprint,
  amsmath,assymb,
 aps,pra,
floatfix,
twocolumn,
superscriptaddress
]{revtex4-2}
\usepackage{graphicx,amsmath}
\usepackage {xcolor}
\usepackage[colorlinks=true, allcolors=blue]{hyperref}
\usepackage[T1]{fontenc}
\usepackage{braket}
\usepackage{bm}
\usepackage[inter-unit-product = \cdot]{siunitx}
\usepackage{dcolumn}
\usepackage{hyperref}
\usepackage{comment}
\usepackage{xcolor}
\usepackage{upgreek}
\usepackage[utf8]{inputenc}
\usepackage[normalem]{ulem}

\newcommand{\am}[1]{\textcolor{black}{#1}}

\begin{document}

\title{Soliton turbulence  of a strongly driven one-dimensional Bose gas}

\author{Manon Ballu}
\affiliation{Université Sorbonne Paris Nord, Laboratoire de Physique des Lasers, 99 av. J.-B. Clément, F‐93430 Villetaneuse, France}
\affiliation{CNRS, UMR 7538, F-93430, Villetaneuse, France}

\author{Romain Dubessy}
\affiliation{Aix-Marseille University, CNRS UMR 7345, PIIM, 13397, Marseille, France}
\author{Aurélien Perrin}
\author{Hélène Perrin}
\affiliation{CNRS, UMR 7538, F-93430, Villetaneuse, France}
\affiliation{Université Sorbonne Paris Nord, Laboratoire de Physique des Lasers, 99 av. J.-B. Clément, F‐93430 Villetaneuse, France}
\author{Anna Minguzzi}
\affiliation{Université Grenoble Alpes, CNRS, LPMMC, 38000 Grenoble, France}

\begin{abstract}
We study the out-of-equilibrium dynamics of a weakly interacting one-dimensional Bose gas in  a box trap, subjected to a drive realized by a periodically oscillating linear potential. After a transient regime, the gas reaches a quasi-steady state,  characterized by the presence of several solitons. At weak driving amplitude, the solitons are only weakly perturbed by one another, while at strong driving amplitude a regime analogous to turbulence is reached, where the solitons are strongly intertwined with each other. We show that a hallmark  of both regimes can be found in  the momentum distribution, which displays  a power-law decay 
$n(k) \sim k^{-2}$ at weak driving amplitude and $n(k) \sim k^{-\alpha}$
with a power-law exponent $\alpha\in [7,9]$ at large amplitude. We further characterize each of the two regimes by following the space-time maps and characterizing the
solitons
using the inverse scattering transform. The protocol analyzed in this study is amenable to experimental realization in current experimental setups.  
\end{abstract}

\maketitle

\section{Introduction}
Turbulence is an ubiquitous phenomenon occurring in viscous non-linear driven classical fluids. It is characterized by an energy cascade, typically  from large to small scales in three spatial dimensions, down to a microscopic scale where dissipation occurs \cite{Barenghi2023,TsubotaBook}. 
Turbulence occurs  as well in quantum fluids, where at zero temperature the defect structures generated by a driving perturbation take the form of quantized vortex lines which may be intertwined among each other \cite{Allen2014}. 
Quantum turbulence was reported in driven superfluid Helium-4 (see e.g~\cite{Skrbek2021}) as well as in Bose-Einstein condensates of ultracold atoms \cite{Henn2009b,Navon2016}. 
In the case of the three-dimensional (3D) weakly interacting Bose gas, a turbulent regime was achieved by driving the gas out of equilibrium by an 
oscillating external potential, either at a frequency close to the radial frequency of a cigar-shaped trap \cite{Henn2009b} or at the resonant frequency of the lowest axial mode in a box trap \cite{Navon2016}, injecting energy at large scale. Due to the intrinsic non-linearity of the system, the energy is transferred to smaller and smaller scales, and a turbulent cascade sets in. It is evidenced by a universal power law in the momentum distribution   $n(k)\propto k^{-3}$, corresponding to a Kolmogorov-Zakharov spectrum \cite{Zakharov_1992}  as predicted by weak-wave turbulence theory \cite{Nazarenko_2011}.  Turbulence was also observed in two-dimensional Bose gases, where dynamical scaling and emergent isotropy was reported \cite{Galka2022}.

In this work, we are interested in the analogous of turbulence in the case of a one-dimensional (1D) weakly interacting Bose gas. This regime is achievable in current state-of-art experiments with ultracold atoms, where specifically it is possible to realize a single 1D tube by trapping the atoms on a magnetic atom chip \cite{Reichel2002,Trebbia2007,Ballu2024} or in an elongated optical dipole trap \cite{Zhang2025}. This allows for high spatial resolution as well as the possibility to apply time-dependent perturbations on the trapped gas \cite{Schuttelkopf2026,Cataldini2025}, conditions necessary to achieve a strongly-out-of equilibrium dynamics  \cite{Dubessy2021} and study the turbulent regime.
From the conceptual point of view, the 1D case is very different from the 3D case, since no vortices can be hosted in one spatial dimension.  \am{ Topological excitations correspond instead to solitons, which can be either bright, characterized by a localized peak in the density for attractive interactions, or dark, characterized by a density dip, whose depth depends on its velocity, for repulsive interactions.} 
Another important difference with respect to the 3D case is that in a  1D weakly interacting Bose gas, 
in absence of an external drive or confinement,  the Gross-Pitaevskii equation describing the system is integrable  by inverse scattering transform, i.e., it has an infinite number of  time-independent constants of motion, that allow the system to be completely described out of equilibrium. Integrable turbulence was studied  in several  equations such as the Korteweg-de Vries equation, the Sine-Gordon equation or the 1D Gross-Pitaevskii equation \cite{Zakharov_2009}.
In the case of the attractive 1D Gross-Pitaevskii equation \cite{randoux_integrable_2016}, 
rogue waves \cite{agafontsev_integrable_2015}\am{, soliton turbulence \cite{dyachenko_soliton_1989} and ballistic correlations \cite{charnay_experimental_2026}  have been reported.  
In the case of repulsive 1D Gross-Pitaevskii equation, relaxation to a generalized Gibbs ensemble has been demonstrated in the context of nonlinear propagation in optical fibers \cite{bastianello_observation_2026}.}
Both for repulsive and attractive interactions, if the gas is subjected to a strong drive, a soliton gas regime is  achieved  \cite{Suret2024}.

We focus here on the 1D repulsive Gross-Pitaevskii equation,
and  we investigate  the effect of the same type of drive as the one employed to generate quantum turbulence in 3D atomic gases \cite{Navon2016}. We show that the drive gives rise to a gas of 
dark solitons, with specific properties depending on the drive strength: at weak drive, a dilute soliton regime occurs, where the solitons are mainly independent,  while at strong drive, we find a dense soliton gas 
regime, 
with intertwined solitons. 
We show that the momentum distribution allows to well identify the two regimes, as they are associated to two different power law scaling: in the dilute soliton regime, we have  $n(k)\propto k^{-2}$, as previously predicted and observed in numerical simulations \cite{Schmidt_non-thermal_2012}, while in the dense soliton regime we find  $n(k)\propto k^{-\alpha}$, with $\alpha \in [7,9]$. 
These two power laws emerge at intermediate momentum scales, while at large momenta, the momentum distribution decreases exponentially with an exponent that can be linked to the momentum distribution of a black soliton.

The paper is organized as follows. In Sec.~\ref{sec:model} we introduce the model Hamiltonian and the drive.  Sec.~\ref{sec:density-maps} shows our results for the space-time density maps in the two regimes. In Sec.~\ref{sec:soliton-count} we introduce a soliton counting method based on the inverse scattering transform, and in Sec.~\ref{sec:momentum-distr} we present our analysis of the momentum distribution in the various soliton gas regimes. Sec.~\ref{sec:exp} discusses the possibilities for experimental observation of our predictions.  Finally, Sec.~\ref{sec:conclusions} contains our conclusive remarks.

\section{Model}
\label{sec:model}
We consider a  gas of $N$ bosons of  mass $m$ confined by a hard-wall box trap  of length $L$ along  $x$.
We assume that the transverse confinement is extremely tight, ensuring  that the condensate remains  in  the transverse ground state all through the dynamics. We refer to Sec.~\ref{sec:exp} below for experimental estimates.  We focus on the case of weak repulsive interparticle contact interactions with one-dimensional coupling strength $g$ and assume that all the particles form a quasi-condensate state, thus neglecting quantum and thermal fluctuations. Quantum corrections to this state at zero temperature scale as  $\sqrt{\gamma}$ with $\gamma= g m/\hbar^2 n_0$  the dimensionless interaction strength and  $n_0=N/L$ the equilibrium density of the gas \footnote{Notice that the density at the center of the trap  is slightly higher than $n_0$ due to the effect of the box trapping.}. 

In this regime of interest, we model the gas  by the time-dependent Gross-Pitaevskii equation for the condensate wave function $\psi(x,t)$:
\begin{equation}
i \hbar \frac{\partial{\psi}}{\partial {{t}}} = - \frac{\hbar^2}{2m} \frac{\partial^2 {{\psi}}}{\partial {{x}^2}} + {g} N |{\psi}|^2 {\psi} + (V_{\rm box} +U ) {\psi}, 
\label{eq:GP_time}
\end{equation}
where $V_{\rm box}$ is the external hard wall  box trapping potential, described  by imposing the boundary conditions $\psi(0,t)=\psi(L,t)=0$, and $U(x,t)$ is the driving perturbation. 
At all times, the condensate wave function is normalized such that $\int dx \, |\psi(x)|^2 =1$. 
For the perturbation, we take a linear potential oscillating sinusoidally with time, according to
\begin{equation}
U(x,t) = \frac{U_{0}}{L}\left(x-\frac{L}{2}\right)\sin(\Omega t). \label{eq:potential}
\end{equation}
This type of drive is the one-dimensional equivalent to what has been used in the 3D experiments  \cite{Navon2016,Galka2022}.
We choose $\Omega = \pi c_0 /L $, with $c_0=\sqrt{\mu_0/m}$ the speed of sound and $\mu_0 = g n_0$ the chemical potential for the system at equilibrium at density $n_0$.
With this choice, the period $T$ of the driving oscillation is equal to the time needed to make a round trip in the box at the speed of sound and allows the system to be excited on a large scale. Note that, 
for small excitation amplitudes, i.e. when $U_{0}< \mu_0$, 
this excitation is resonant with  the lowest phonon excitation in the box trap,  estimated in the limit where the healing  length $\xi_0=\hbar/\sqrt{2gn_0m}$ is much smaller than the size  $L$ of the box. In the linear response regime, holding in the weak amplitude limit, the dynamical evolution of the gas is well accounted for by phonon excitations.  In our study, we are mainly interested in understanding  the strong amplitude excitation regime beyond linear response, where multi-phonon excitations can take place and the creation of topological excitations, such as solitons, may occur.

All the results presented in the forthcoming sections are obtained by solving numerically the time-dependent Gross-Pitaevskii equation (\ref{eq:GP_time}) with a high-precision algorithm allowing to account for strongly varying features in space. This requires  a short time-step for the temporal evolution  (see details  in  Appendix~\ref{app:numerical-method}) as well as to project out large-momentum modes to avoid aliasing problems.

\section{Density maps}
\label{sec:density-maps}

\begin{figure}[h]
\begin{center}
    \includegraphics{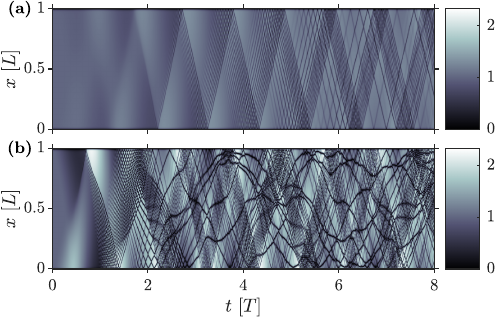}
    \caption{{\label{fig:densite}} Density profile maps $n(x,t)=N |\psi(x,t)|^2$ in units of the mean density $n_0=N/L$. Space coordinate $x$ is expressed in units of $L$ while time $t$ is in units of $T=2L /c_0$, corresponding to  the time it takes to make a round trip in the box at the speed of sound. Interaction strength fulfills $\tilde{g}N=10^4$ with $\tilde{g}=gmL/\hbar^2$. The driving amplitude is set to \textbf{(a)} $U_0= 0.1\, \mu_0$ and \textbf{(b)} $U_0=\mu_0$.
    }
    \end{center}
\end{figure}

We first analyze the density profile $n(x,t)=N|\psi(x,t)|^2$ in space and  time during the excitation by the external driving potential $U(x,t)$.
In Fig.~\ref{fig:densite}, we show the density maps in the space-time plane, for two cases of excitation amplitudes,   $U_0 = 0.1 \mu_0$ and $U_0 = \mu_0$ corresponding to a weak and a strong excitation, respectively. 
In both cases,  the space-time density maps show  an oscillation of the density along the box, with alternating light areas corresponding to a density above the mean density $n_0$, and dark areas corresponding to a density below the mean density.  The maps also show the appearance of trains of 
dark solitons, corresponding to a local density depletion.

For the case of a weak amplitude of the driving potential, see Fig.~\ref{fig:densite}~(a), the solitons are  well separated, and propagate at speeds close to the speed of sound, in a straight line, with reflections off the edges of the box. It is visible on the figure that, in trains of solitons with almost parallel trajectories, the right-most solitons ---i.e., the ones moving more slowly--- are more contrasted than the left-most solitons. This is expected from the relation between the propagation velocity and the density dip of a 
dark soliton, which can be deduced from the solution for the wave function of a single soliton, here given in the 
thermodynamic 
limit (with $L\to\infty$ at fixed $n_0$):
\begin{align}
\psi(x,t) &= \frac{ e^{- i \mu_0 t}}{\sqrt{L}} \left[ \sqrt{1-\nu^2} \tanh \left(\frac{[x-x_s(t)]\sqrt{1-\nu^2}}{\sqrt{2} \xi_0}\right)\right.\notag\\
 & \qquad \qquad \qquad \qquad \qquad \qquad  \left. + i \nu \right].
\label{eq:fonction_soliton}
\end{align}
Here, $\nu=v/c_0$ 
is the speed of the soliton divided by the speed of sound, $x_s(t)=x_0+ v t$ is the soliton position.

At higher driving amplitude, see Fig.~\ref{fig:densite}~(b),  more numerous and narrower solitons appear, with greater density contrast.  Over several  oscillation periods, the system turns into a state with several intertwined 
solitons, difficult to isolate, forming tangles in space time.  This state  appears to be a (quasi-)steady state, as obtained by inspecting the  density at a fixed spatial point as a function of time.
The velocities of the solitons are more diverse than in the case of weak excitation amplitude:
the velocity of a given soliton is not constant, 
and some very dark and slow solitons even change the sign of their velocity and do not reach the edge of the box.
We can also see from the slopes of the trajectories that some light solitons  move faster than the equilibrium speed of sound $c_0$.
This means that
in these very non-equilibrium systems, the effective speed of sound is larger than  $c_0$. 
In the very dilute limit where the $N_s$ solitons present in the system do not overlap, we can describe the total wave function as a product of $N_s$ independent solitonic wave functions of the form Eq.~\eqref{eq:fonction_soliton} \cite{Schmidt_non-thermal_2012}, such that the total volume occupied by $N_s$ solitons scales linearly with $N_s$. We can then interpret the increase of the speed of sound by an increase of the background atomic density due to an exclusion-volume effect: the background density is of order $N/(L-N_s\ell_s)=n_0/(1-n_s\ell_s)$ where $n_s=N_s/L$ is the soliton density and $\ell_s$, the average volume excluded by a single soliton, is of the order of the healing length. This results in a modified, larger sound velocity 
$c_0/\sqrt{1-n_s\ell_s}>c_0$. 
This  also yields an effective, reduced healing length 
$\xi_{\rm eff}=\xi_0\sqrt{1-n_s\ell_s}<\xi_0$. A further analysis in support to this description is provided in  Appendix \ref{app:effective_xi} and Fig.~\ref{fig:ls_over_xiLax}.

\section{Soliton counting method}
\label{sec:soliton-count}
Since one of the main features in the density profiles at strong excitation drive is the presence of several solitons, and we wish to characterize the state of the system through the estimate of their number and speed, we specifically implement a procedure of soliton counting based on the inverse scattering transform and the Lax spectrum~\cite{RomainSciPost}. This method allows to extract the number of solitons and their speeds in an easy and reliable way, in contrast to the mere inspection of the density profile, which is especially challenging to analyze in the dense soliton regime.  

The inverse scattering transform is  a mathematical method to solve nonlinear partial differential equations 
(see e.g. \cite{Ablowitz_solitons_1981} for the  general formalism), which applies to a family of  integrable equations, and in particular to the Gross-Pitaevskii equation in the absence of an external potential \cite{Zakharov_1972,Faddeev2007Hamiltonian}.
According to this method, the solution of the Gross-Pitaevskii equation is found by solving an inverse scattering problem: a suitable Lax operator $\mathcal{L}$ is defined, in which the condensate wave function $\psi$ enters as a potential, and whose expression is obtained 
by the knowledge of the scattering amplitudes.
The Lax operator for the Gross-Pitaevskii equation  reads 
\begin{equation}
    \mathcal{L}=\frac{i\hbar }{2 m}\begin{pmatrix}
\displaystyle\frac{\partial}{\partial x}&-\displaystyle\sqrt{N k_c } \psi\\
\sqrt{N  k_c} \displaystyle \psi^*&-\displaystyle\frac{\partial}{\partial x}\\
\end{pmatrix},
\end{equation}
with $k_c=m g/\hbar^2$.
The eigenvalues of the Lax operator, which have the dimension of a velocity, can be shown to be constants of motion during the dynamics. In the thermodynamic limit, corresponding to sending the system size $L \rightarrow \infty$ while keeping the density $n_0$ constant, the eigenvalues $\nu_j$  can be divided into two groups: a continuous part, corresponding to dispersive solutions and consisting of two branches, separated by a gap where we find  a discrete set of eigenvalues, each of which is associated to a soliton \cite{Faddeev2007Hamiltonian}. Hence, the number of discrete eigenvalues of the Lax operator $\mathcal{L}$ can be used to count the number of solitons and identify their speed $v$, which is directly related to the eigenvalue $\zeta$ through $v=-2\zeta$.

We apply this idea to estimate at a given time $t$ during the dynamics the number of solitons in the system: we input in the Lax operator the condensate wave function $\psi(x,t)$ and check for the number of discrete eigenvalues of the Lax operator.

\begin{figure*}[t]
\begin{center}
  \includegraphics{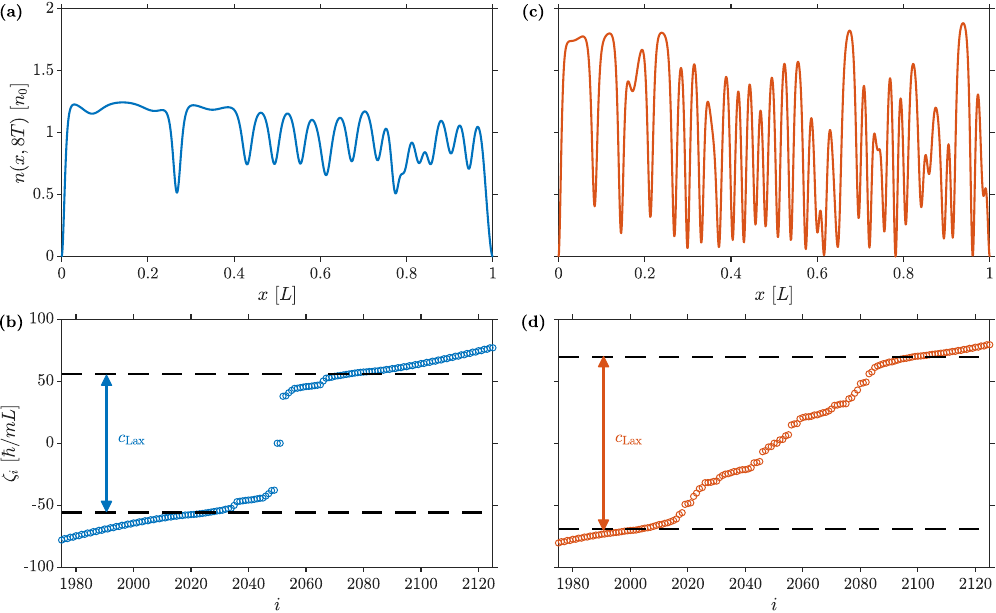}
     \end{center}
    \caption{ \label{fig:lax_spectrum} Density profile in units of $n_0$ at $t=8T$ for \textbf{(a)} $U_0=0.1\mu_0$ and \textbf{(c)} $U_0=1\mu_0$. The central part of their respective Lax spectrum is shown in \textbf{(b)} and \textbf{(d)}, with the eigenvalue $\zeta_i$ as a function of the index $i$. The dashed black lines delimits the inner region of the spectrum where eigenvalues correspond to solitons. Details on the method to extract this information are given in the main text and in Appendix~\ref{app:counting}. 
    }
\end{figure*}

In our case, a few subtleties need to be addressed.
Firstly, the equation we use is not integrable because of the driving potential $U$. However, the potential vanishes at each half period: $U(x,pT/2)=0$ with $p$ an integer. If we stopped the driving at this specific time, the equation would become integrable and the Lax spectrum would be conserved from this time on. In the data presented in this paper, we thus applied the method at times $t=8T$ and $t=32T$, both verifying the previous condition.
Secondly, the system is not uniform since it is trapped by an external box trap. Still, it is possible to extrapolate to the thermodynamic limit by imposing effectively periodic boundary conditions to the system. This is achieved by doubling the size of the system and antisymmetrizing the wave function. The number of solitons in this new system of size $2L$ is doubled compared to the initial one: for each soliton present in the initial system, there is a twin soliton with opposite velocity in the doubled system \footnote{Note that the identification of the sign of the soliton velocity in the initial box of size $L$ is not straightforward. It would require a careful identification of the position of the solitons from the eigenfunction of the Lax spectrum, and an examination of the variation of the phase of the wave function $\psi$ at this position. Here we are mostly interested in the modulus of the velocity and we do not look at its sign.}.
Finally, another difficulty remains: because of the finite size of the system, all the eigenvalues are discrete, making it very hard to identify the ones associated to solitons. To overcome this issue we adapt the method developed in \cite{RomainSciPost}. We double the size of the system a second time using the periodic boundary conditions, realizing a simple copy of the system, see Appendix~\ref{app:counting}. The number of solitons in the final system of size $4L$ is again doubled compared to the intermediate system of size $2L$, and each soliton is present twice, with the same eigenvalue in the Lax spectrum, whereas for the (quasi) continuous branches this transform results in a denser spectrum, without additional degeneracy. To identify the solitons, we can thus compare the degeneracy of each eigenstate between the systems of size $2L$ and $4L$: if the degeneracy of a given eigenvalue is doubled, it corresponds to a soliton (see details in Appendix~\ref{app:counting}). 

In addition to counting the number of solitons, the Lax spectrum allows us to extract their velocities \cite{Note2}.
We also deduce the effective speed of sound $c_\textrm{Lax}$ of the system from the two gap edges of the (quasi) continuous branches $\zeta^{\pm}$, with
$c_\textrm{Lax} = \zeta^+ - \zeta^-$ \cite{RomainSciPost}.

The result of the diagonalization of the Lax operator is shown in Fig. \ref{fig:lax_spectrum} for two typical cases, weak excitation $U_0=0.1\mu_0$ for the left graphs, (a) and (b), and stronger excitation $U_0=\mu_0$ for the right graphs, (c) and (d). The upper graphs, (a) and (c), show the density, while the eigenvalues are presented in the lower graphs, (b) and (d), for the intermediate system of size $2L$. The solitons lie in the central window of size $c_{\rm Lax}$ between the dashed lines. Half of these solitons, which can have positive or negative velocities, are present in the original system of size $L$ represented in the upper panel, the other half with opposite velocity \cite{Note2} arise from the symmetric system. The two central eigenvalues $\zeta=0$ correspond to the two zeros of the density due to the box boundary condition at position $0$ and $L$ of the periodic system of size $2L$, see Fig.~\ref{fig:doubling2}.

For a weakly excited system, shown  in Fig.~\ref{fig:lax_spectrum} (a)-(b), the Lax spectrum consists mainly in shallow solitons, with velocities close to the speed of sound (except for the two $\zeta=0$ eigenvalues).  $N_s=23$ solitons are found within
the initial system of size $L$. 
By inspecting the density profile in Fig.~\ref{fig:lax_spectrum}(a), it appears clearly 
 that solitons are present,  displayed by
 a dozen marked dips. 
 However, 
 very shallow solitons close to the gap edges cannot be easily spotted from the simple examination of the density,
which  
highlights
the usefulness of the counting method.

For a strongly excited system, shown in Fig.~\ref{fig:lax_spectrum}(c)-(d), the Lax spectrum contains more solitons,  including low-velocity solitons, with a high contrast. Specifically, we obtain $N_s=46$ solitons in the system of size $L$. The distribution of the soliton eigenvalues in the gap tends to be more uniform. 

The comparison of the weak and strong excitation cases reveals that the effective speed of sound $c_{\rm Lax}$, determined from the edges of the continuous branches of the Lax spectrum, increases with the number of solitons. The associated healing length, $\xi_{\rm Lax}=\hbar/(\sqrt{2}mc_{\rm Lax})$, decreases with the number of solitons. This can be intuitively understood in the finite size box with fixed total atom number through the exclusion-volume effect mentioned above: the effective background density increases because of the density depletion of each soliton, with a higher contribution for slower (deeper) solitons (see also Appendix~\ref{app:effective_xi}).

\section{Momentum distribution}
\label{sec:momentum-distr}
In experiments, in situ observation of solitons is challenging due to the required imaging resolution. In this section, we show that the momentum distribution also contains clear information on the soliton distribution generated by the driving potential.
The momentum distribution is obtained from 
$\hat \psi(k,t)$,
the Fourier transform of the condensate wave function
according to  $n(k,t)= |\hat \psi(k,t)|^2$  (see Appendix \ref{app:numerical-method} for details).
The momentum distribution is further averaged over 6 periods at the end of the time evolution, well after the time required for the system to reach its steady state, in order 
to access the ensemble average of the state:
$\bar n(k)=\frac{1}{6T} \int_{26T}^{32T} dt \,   n(k,t)$.

\begin{figure}[htb]
\begin{center}
    \includegraphics{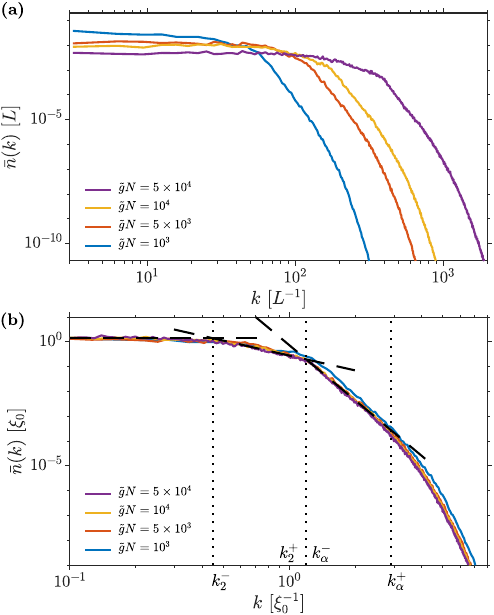}
    \caption{{\label{fig:gN}}
    \textbf{(a)} Time averaged momentum distribution  $\bar n(k)$ in units of $L$ 
    for $U_0=\mu_0$ and various values of the particle 
    number multiplied by the dimensionless interaction strength $\tilde{g}N$. \textbf{(b)} All curves collapse on a single one by a proper choice of units involving their respective healing length $\xi_0$. The dashed lines indicate the power law behaviors $k^{-2}$ and $k^{-7.4}$. 
    }
    \end{center}
\end{figure}

We first analyze the behavior of the momentum distribution for various values of the dimensionless  parameter $\tilde g N$, where $\tilde g = m g L /\hbar^2$, which characterizes the strength of the interaction in the Gross-Pitaevskii equation. The time-averaged momentum distribution is shown in Fig.~\ref{fig:gN} for an intermediate value of the driving amplitude  $U_0=\mu_0$. Specifically, we focus on the tails of the momentum distribution, and hence represent it in log-log scale.
We find that all curves collapse on a single one if we rescale both axes by the healing length $\xi_0$, thus showing that solitons, whose size is fixed by the healing length, play a major role in the large-$k$ behavior of the  momentum distribution. 

\begin{figure*}[ht]
\begin{center}
   \includegraphics{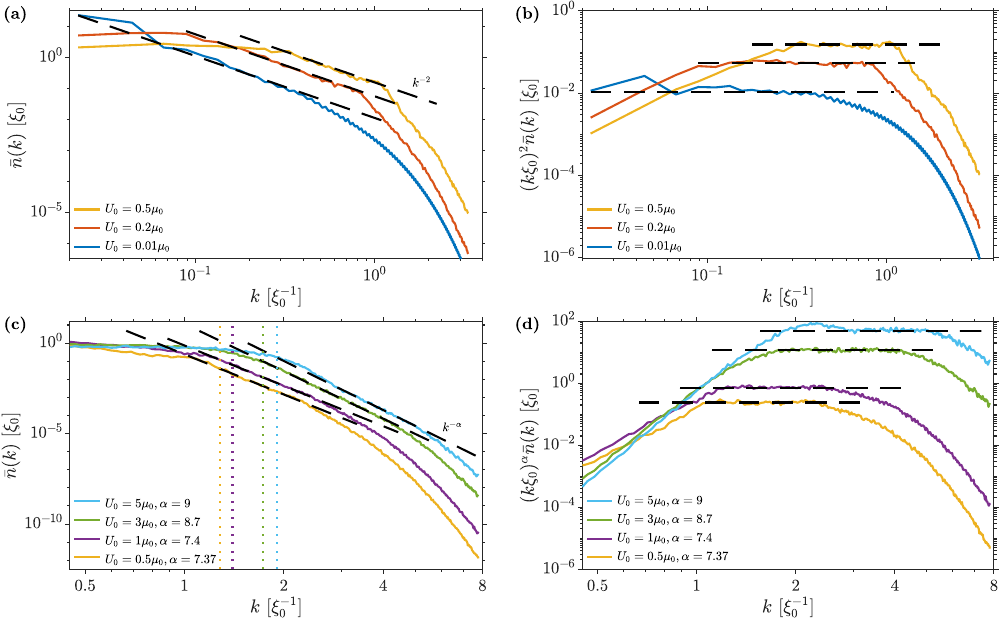}
   \caption{{\label{fig:nk_allU}}  Time averaged momentum distribution $\bar n(k)$ in units of \am{$\xi_0$} as a function of the wave vector $k$ in units of $\xi_0^{-1}$ in double logarithmic scale, for weak driving amplitudes \textbf{(a)} and large driving amplitudes \textbf{(c)}. For all curves, the power-law behavior is highlighted by the  dashed black lines:  $\bar n(k) \sim k^{-2}$ in \textbf{(a)}, and $\bar n(k) \sim k^{-\alpha}$  in \textbf{(c)}  with exponent $\alpha=$ $7.37$ (yellow), $7.4$ (purple), $8.7$ (green) and $9$ (blue), from bottom to top. The vertical dotted lines in \textbf{(c)} indicate the corresponding value of $\xi_{\rm Lax}^{-1}$ for reference. The power-law behaviour is highlighted in panels \textbf{(b)} and  \textbf{(d)}  by displaying $k^2\bar{n}(k)$ (respectively  $k^{\alpha} \bar n(k)$) in log-log scale for the same driving amplitude parameters $U_0$ as in \textbf{(a)} (respectively \textbf{(c)}). All curves were obtained for $\tilde{g}N=10^4$ and averaged over the last six oscillation periods.}
       \end{center}
  \end{figure*}

We then study the momentum distribution for various strengths of the driving field $U_0$.
At weak driving amplitude $U_0 \leq \mu_0$, as shown in Fig.~\ref{fig:nk_allU}, top panels, we clearly identify  a  power-law decay $\bar n(k)\sim k^{-2}$ at intermediate wave vectors, between the limits labeled $k_2^-$ and $k_2^+$ in Fig.~\ref{fig:gN}.  \am{As demonstrated in Ref.~\cite{Schmidt_non-thermal_2012}, in the dilute soliton regime one can estimate the momentum distribution by assuming that the condensate wave function is approximately described by a gas of randomly centered dark solitons. By averaging over the random positions of the soliton centers and Fourier transforming,   }
the average momentum distribution at weak driving amplitude \am{takes the form}
\begin{equation}
\label{eq:nk_dilute}
  \bar n(k)\simeq \frac{4 n_\textrm{s} n_0 \eta}{4 n_\textrm{s}^2\eta^2+k^2}\left[\frac{k /k_+}{\sinh\left(k /k_+ \right)}\right]^2.
\end{equation}
Here, $\eta=1-\langle \nu^2\rangle$ is related to the average over the relative velocities $\nu=v/c_{\rm Lax}$ of all the solitons extracted from the Lax spectrum at $t=32T$, from which we also extract the soliton density $n_s=N_s/L$. The 
wavevector scale $k_+$ 
corresponds to the typical scale 
of the large-$k$ exponential decay of the momentum distribution and \am{we obtain it}  by fitting the numerical results with an exponential tail.
In the case of a single black soliton, we expect $k_+=\sqrt{2}/(\pi\xi_0)$ to match the soliton wave function in momentum space, which scales as $1/\sinh(\pi\xi_0 k/\sqrt{2})$. 

\begin{figure}[ht]
\begin{center}
    \includegraphics{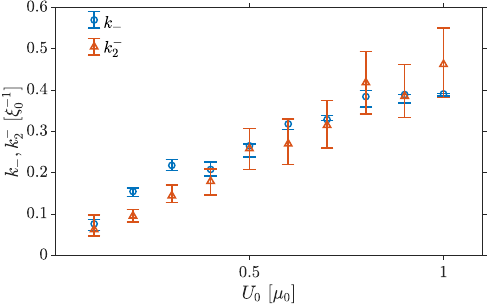}
    \caption{{\label{fig:debut_k2}}
    Wavevector  $k_2^-$ corresponding to the onset of the $k^{-2}$ power law decay of the momentum distribution in the simulations as compared to the estimate $k_-=2n_s[1-\langle\nu^2\rangle]$, both in units of $\xi_0^{-1}$, as a function of the driving amplitude $U_0$ in units of $\mu_0$. See text for detail, and Appendix~\ref{app:counting} for the estimate of the uncertainty.}
\end{center}
\end{figure}

Equation (\ref{eq:nk_dilute}) indeed predicts an algebraic decay  $\bar n(k)\sim k^{-2}$  between two boundaries, $k_-=2n_s\eta=2n_s[1-\langle\nu^2\rangle]$, related to the distribution of solitons, and $k_+$.
We have tested this prediction on our numerical data using the soliton counting method described in Sec.~\ref{sec:soliton-count}: Fig.~\ref{fig:debut_k2} 
compares $k_-$ to the lower bound $k_2^-$ of the region in momentum where we find that the $k^{-2}$ scaling holds, as determined from the intersection in a log-log plot from linear fits with slopes 0 and -2, see Fig.~\ref{fig:gN}(b). 
We find a good agreement between these two quantities, which means that measuring the momentum distribution could be a way to estimate the soliton density in an experiment. 

At larger amplitude of the drive, for $U_0\geq 0.3\,\mu_0$, we observe the emergence of a second power-law regime for larger momenta in the interval $k_\alpha^-<k<k_\alpha^+$, with $k_\alpha^-\simeq\xi_{\rm Lax}^{-1}$, as shown in Fig.~\ref{fig:nk_allU} (c).
This latter  power-law exponent is much larger, of the order of  $\alpha\simeq 8 $
and corresponds to the case of several intertwined solitons, where Eq.~\eqref{eq:nk_dilute} breaks down.
This exponent is also higher than those usually found in Kolmogorov-Zakharov spectra \cite{Nazarenko_2011}.
At present, we have no theoretical prediction  for this power law. We note that similar large power exponents have been observed in simulations for polariton systems for regimes dominated by defects and solitons \cite{He_2017,Vercesi_2023}.
We finally note that at intermediate values of the drive, for $U_0 \in [0.3 \mu_0, \mu_0]$, we find coexistence of the two power law regimes in the momentum distribution \am{(see again Fig.~\ref{fig:gN})}. In this regime indeed in the space-time maps we find both weakly excited dilute solitons and more intertwined ones. 

\begin{figure}[ht]
\begin{center}
    \includegraphics{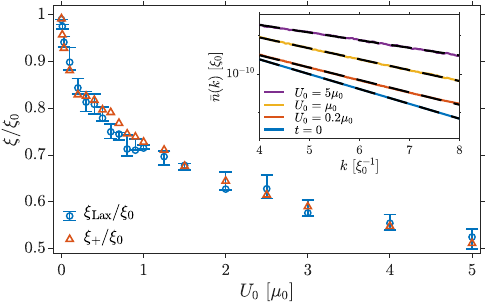}
    \caption{{\label{fig:queue_dis_xieff}}
Comparison of the ratio $\xi_\textrm{Lax}/\xi_0$  as obtained from the Lax spectrum and $\xi_+/\xi_0$ obtained by fitting the time-averaged momentum distribution $\bar{n}(k)$ for $k \in [4\xi_0^{-1}, 8\xi_0^{-1}]$ with the function $\exp(-\pi \sqrt{2}\xi_+ k)$, as a function of the driving amplitude $U_0$, in units of $\mu_0$. See Appendix~\ref{app:counting} for the estimate of the uncertainty on $\xi_{\rm Lax}$. The inset shows the tail of the time-averaged momentum distribution $\bar{n}(k)$ in units of \am{$\xi_0$} in semi-logarithmic scale for various values of the driving amplitude $U_0$, indicated in the legend,  as a function of the wave vector $k$ in units of $\xi_0^{-1}$. The black dashed lines correspond to the results of the fit used to extract $\xi_+$ in the main panel. In the inset we also show for comparison the initial momentum distribution $n(k, t=0)$.
In this case, the exponential tail is related to the sharp decrease of the density at the two edges of the box.
All data were obtained for $\tilde{g}N=10^4$ and the momentum distributions were averaged over the last six oscillation periods.}     
    \end{center}
\end{figure}

At large wave vectors $k>k_\alpha^+$, the momentum distribution shows an exponential decay. For a system with a single black soliton, the momentum distribution scales as  $n(k) \sim \exp(-\pi \sqrt{2} \xi_0 k)$. Inspired by the prediction of Eq.~\eqref{eq:nk_dilute}, we fit the exponential decay for  all amplitudes $U_0$, including in the large amplitude regime, with an exponential law 
$n(k) \sim  \exp(-2 k/k_+)$, and determine the adjustable parameter $k_+$, to which we associate a healing length $\xi_+=\sqrt{2} ( \pi k_+)^{-1}$. 
We find that $\xi_+$ is lower than $\xi_0$ and decreases as $U_0$ increases. This is another indication of a renormalization of the healing length due to the presence of many solitons.
Figure~\ref{fig:queue_dis_xieff} shows a very good agreement at all drive amplitudes $U_0$ between the length $\xi_+$ extracted from the fit of the momentum distribution at large $k$, and $\xi_\textrm{Lax}$ as deduced from the Lax spectrum.  
Hence, the momentum distribution at large momenta is determined by the short-distance behavior of the fluid at the scale of each single soliton, with renormalized sound velocity corresponding to the compressibility of the soliton gas regime. The data compare also well with the estimate $\xi_{\rm eff}=\xi_0\sqrt{1-n_s\ell_s}$
if we compute $\ell_s$ as the average volume occupied by a soliton in the system, see Appendix~\ref{app:effective_xi} and Fig.~\ref{fig:ls_over_xiLax}.

Finally, we have verified that the emergence of power-law tails in the momentum distribution is not specific to the type of excitation drive, and similar results are obtained by driving  the system through  a moving Gaussian potential (see Appendix \ref{app:other-drive}).

\section{Discussion on experimental feasibility}
\label{sec:exp}

In this section, we discuss the possibility of an experimental implementation of the model presented in the previous sections. Box trapping potentials are routinely created in ultracold-atom experiments~\cite{vanEs2010,Tajik2019,Navon2021}. To ensure the 1D nature of the system, it is necessary to check that the chemical potential $\mu\simeq \mu_0$ is small compared to the energy associated with the transverse confinement. For a transverse harmonic trap of frequency $\omega_\perp/2\pi$, the 1D coupling strength is $g\simeq2\hbar\omega_\perp a_s$, where $a_s$ is the atom scattering length \cite{Olshanii1998}. In the following we focus on $^{23}$Na atoms in the $|F=1,m_F=-1\rangle$ internal state
with $a_s=54.54a_0$~\cite{Knoop2011}, 
where $a_0$ is the Bohr radius, $F$ is the total angular momentum of the atom, and $m_F$ its projection along the quantization axis.
As realistic experimental parameters, we consider $\omega_\perp/2\pi=\SI{4}{\kilo \hertz}$, $L=\SI{140}{\micro \meter}$ and $N=1500$. This leads to $\mu\simeq h \times \SI{247}{\hertz}$, yielding $\xi_0\simeq 1\,\SI{}{\micro \meter}$, allowing in situ observation of the solitons with a high-aperture objective \cite{Cataldini2025},
$c_0 \simeq 2$\,mm.s$^{-1}$ and $T=2L/c_0\simeq\SI{140}{\milli \second}$.

The excitation potential $U(x,t)$ can be created by a magnetic field gradient induced by a pair of coils of axis $x$ in anti-Helmholtz configuration, carrying a sinusoidal current of period $T$. In order to obtain an amplitude $U_0 \simeq \mu_0$, the magnetic field gradient should reach \SI{25}{\milli G \per\centi\metre}, which is easily achievable.

In order to measure the momentum distribution, it is possible to rely on the focusing technique \cite{Shvarchuck2002,vanAmerongen2008}. It enables to map the momentum distribution of the system at time $t$, $n(k,t)$, onto its density distribution $n_f(x)$ after the following evolution: 
first a longitudinal harmonic potential $V_\textrm{harm}(x)=m\omega_x^2x^2/2$ is suddenly applied for a duration $\tau$; second, all confinements are switched off and the gas expands for a duration $t_f$. In practice, $\omega_x\tau \ll 1$ so that the atoms can be considered static while the harmonic potential is on. Moreover, the switch-off of the transverse trapping leads to a decrease of interaction energy over a timescale $\omega_\perp^{-1}$. This timescale is short enough to neglect the interactions during expansion if the longitudinal energy is smaller than $\hbar\omega_\perp$. In the one-dimensional limit, corresponding to $\hbar\omega_\perp\gg\mu$, a longitudinal energy of order $\hbar\omega_\perp$ would thus be dominated by free particles and  yield a maximum wave vector $k\lesssim\xi^{-1}\sqrt{\hbar\omega_\perp/\mu}=\xi^{-1}/\sqrt{2n_0a_s}$, where the bulk healing length $\xi$ is well estimated by $\xi_{\rm Lax}$. To explore the momentum distribution up to $k_\alpha^+$ 
which reaches a few $\xi_0^{-1}$, the density should thus be significantly smaller than $1/(2a_s)$ which for sodium is \SI{170}{\per\micro\metre}. Considering the typical values $N=1500$ for the atom number and $L=\SI{140}{\micro\metre}$ for the box length, the above requirement for the density dilution is easily fulfilled.

Finally, we stress that even if the spatial resolution is not sufficient to resolve the individual solitons, in the weak drive regime the observation of the 
$k^{-2}$
power-law decay of the momentum distribution would provide a strong indication of the presence of solitons.
Assuming $\omega_x/2\pi=\SI{100}{\hertz}$, $\tau=\SI{600}{\micro\second}$ and
$t_f\simeq\left(\omega_x^2\tau\right)^{-1}\approx\SI{4}{\milli\second}$~\cite{jacqmin_momentum_2012-1}, the typical peak density of the momentum distribution $n_f(0)$ is of the order of $\SI{56}{\per\micro\meter}$  and $\xi_0^{-1}$ corresponds after expansion to a distance close to $\SI{13}{\micro\meter}$.
At low modulation amplitude, observing the $k^{-2}$ power law up to $k_2^+$ requires a sensitivity below $10^{-2}n_f(0)$. The onset of the power law around $k_2^-$ is accessible with standard absorption imaging techniques \cite{Smith2011,jacqmin_momentum_2012-1,Tauschinsky2013}, while accessing the full power law up to $k_2^+$ may require fluorescence imaging with single atom sensitivity \cite{Bcker2009,Bergschneider2018}.
A similar sensitivity is needed for observing at least part of the $k^{-\alpha}$ power law for large modulation amplitude.
Observing this power law up to $k_\alpha^+$ in the intermediate modulation regime ($U_0\sim0.5\mu_0$) and above is a real experimental challenge since a sensitivity up to $\sim10^{-4}n_f(0)$ is needed. Single atom sensitivity (fluorescence or electronic detection~\cite{Chang2016}) and averaging over several thousands repetitions would be a minimum requirement.

\section{Conclusions and outlook}
\label{sec:conclusions}
In conclusion, we have studied the effect of a periodic driving potential on a weakly interacting Bose gas. Depending on the driving amplitude, we have found two very different regimes, both characterized by a power law decay in the momentum distribution. At weak driving amplitude, we identify few, isolated 
dark solitons essentially propagating in an independent manner. In this regime, 
the momentum distribution
follows a power law $n(k)\sim k^{-2}$ as predicted by the independent soliton model \cite{Schmidt_non-thermal_2012}. We have checked that the lower bound $k_2^-\simeq k_-$ of the power-law region is related to the soliton density $n_s$, which we have estimated using the Lax method, 
while the tail of the distribution is exponential, decaying  on a scale related to the size of the soliton core. At intermediate drives, we observe the onset of a second power-law regime at larger wave vectors, with a larger exponent  $\alpha\sim 8$.  This second regime fully emerges at strong driving amplitude and corresponds to  soliton turbulence where several 
dark solitons are intertwined in the space-time maps. 
We have verified that the power law and its exponent at strong drive are very general and independent from the type of driving potential, and hence may be considered as a hallmark of the soliton-turbulence regime.
Such a power law decay with a large exponent was also  observed in other instances of non linear Schr{\"o}dinger equation, when several solitons are present \cite{He_2017,Vercesi_2023}, though in a different context, where the Bose gas is subjected to pump and losses.

Our study has been performed at zero temperature, and we have neglected particle losses. In realistic experimental conditions,
atoms may be lost from the trap due to the finite height of the box edges, or due to collisional processes.  Furthermore, in the experiment, thermal fluctuations  could induce additional heating.
As an outlook, it could be interesting to extend this work to account both for particle losses and finite temperature effects.
Furthermore, our study shows that 
it would be very interesting to be able to access in situ density profiles, in particular to observe the appearance of solitons in the system, in the spirit of the in situ observation of quantum vortices \cite{Wilson2015}. It remains an open experimental challenge to develop an imaging method with a resolution lower than the healing length.
Finally, theoretically, it would be 
interesting to address the nature of the state of the system at long times, and whether it is thermalized or rather non-thermal \cite{Erne2018}.

\acknowledgements
We thank M. Olshanii and L. Canet for very stimulating discussions.  
LPL is UMR 7538 of CNRS and Université Sorbonne Paris Nord. LPMMC is UMR 5493 of Université Grenoble Alpes and CNRS.
We acknowledge funding from the ANR Quantum-SOPHA project ANR-21-CE47-0009 and from the France 2030 `Dyn1D' project ANR-23-PETQ-0001.
R.D. acknowledges support from the French government under the France 2030 investment plan, as part of the Initiative d'Excellence d'Aix-Marseille Université -- AMIDEX AMX-22-CEI-069.

\section*{Data availability}
The data that support the findings of this article are openly
available \footnote{M. Ballu, R. Dubessy, A. Perrin, H. Perrin, and A. Minguzzi, Data set for ``Soliton turbulence of a strongly driven one-dimensional Bose gas'' [Data
set], Zenodo (2026), http://doi.org/10.5281/zenodo.21072102}.

\appendix

\section{Numerical method}
\label{app:numerical-method}
 We provide details in this section about the numerical method used to solve the time-dependent Gross-Pitaevskii equation (\ref{eq:GP_time}). 
To simplify notations, we will  use dimensionless units throughout this section, with lengths in units of $L$, energies in units of $\hbar^2/(mL^2)$ and time in units of $mL^2/\hbar$.
We use a spectral method especially tailored for
 box-like potentials. It relies
 on the discrete sine transform (DST), which,  using
$N_p$ grid points, reads:
\begin{equation}
\hat \psi_k=\sum_{j=1}^{N_\textrm{p}}\psi_j\sin\left(\frac{\pi k j}{N_\textrm{p}+1} \right)
\end{equation}
for a function $\psi_j=\psi(x_j)$ defined in dimensionless units at positions $x_j=j/(N_p+1)$ for $j \in [\![1;N_\textrm{p} ]\!] $ (points $0$ and $1$ are excluded),
and the inverse sine transform (DST$^{-1}$), given by 
\begin{equation}
\psi_j=\frac{2}{N_\textrm{p}+1}\sum_{k=1}^{N_\textrm{p}}\hat \psi_k\sin\left(\frac{\pi k j}{N_\textrm{p}+1} \right).
\end{equation}
Notice that  we use the discrete sine transform rather than the Fourier transform, as this allows to enforce the hard-wall condition $\psi(0)=\psi(1)=0$ at each time step.

In order to study the time evolution under periodic drive, we  first initialize the wave function to the ground state in the box trap potential. This is achieved using evolution in  imaginary time, setting to  zero the external driving  potential, while keeping the normalization constant in the numerical code to
$1=\sum_j |\psi_j|^2$.
After equilibration, the condensate wave function is then evolved by solving the time-dependent Gross-Pitaevskii equation in real time with $U(x,t)$ as the external potential.

In order to avoid spectrum  aliasing problems
in the solution of the Gross-Pitaevskii equation,
we restrict the $k$ grid to a cutoff value $k_\textrm{cut} = 2 k_\textrm{max}/3$, with $k_{\rm max}=\pi N_p$.
Let $P$ be the projection onto this restricted $k$ space. The GPE is implemented as:
\begin{equation}i\frac{d\hat{\psi}}{dt} = P\left( \frac{1}{2}k^2 \hat{\psi}-\tilde \mu \hat{\psi} + \textrm{DST} \left[ \rho \psi \right]\right)
\end{equation}
with
\begin{equation}
\rho=\textrm{DST}^{-1} \left\{ P \left[\textrm{DST} \left( \tilde gN|\psi|^2+\tilde U \right)\right] \right\}.
\end{equation}

The dimensionless quantity are labeled by a tilde. We have run simulations testing different values of the excitation amplitude of the scaled potential $\tilde{U}_{0}=U_0mL^2/\hbar^2$, which is relevant to compare with the equilibrium chemical potential $\tilde{\mu}$. Furthermore, we have taken $\tilde{g} N= 10^4$ for most simulations, as this is the order of magnitude that would be achievable on our cold atom experiment, see Sec.~\ref{sec:exp}.
We have taken a uniform spatial grid with $N_\textrm{p}= 1024$ points. We thus have a space between two points $\Delta x=1/(N_\textrm{p}+1)=\SI{9.8e-4}{}$ small compared to $\xi_0=1/\sqrt{2\tilde{g}N}=\SI{7e-3}{}$ for the scaled quantities. Finally, we have checked that the results  do not depend on the specific choice of the momentum cutoff $k_\textrm{cut}$.

\section{Result for alternative drive mode}
\label{app:other-drive}

\begin{figure}[htb]
\begin{center}
    \includegraphics{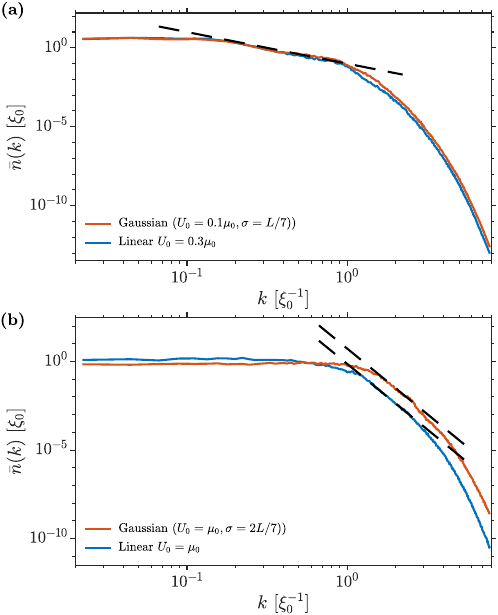}
    \caption{{\label{fig:other-drive}} 
    \textbf{(a)} Comparison of the time-averaged momentum distribution $\bar{n}(k)$ in units of \am{$\xi_0$} in log-log scale as a function of the wave vector $k$ in units of $\xi_0^{-1}$ for two excitation schemes: in blue relying on the linear driving potential $U$ while in red with the Gaussian repulsive potential $U_\sigma$. The black dashed line corresponds to a $k^{-2}$ power law fit. \textbf{(b)} Same for larger excitation amplitudes. The black dashed lines correspond to $k^{-\alpha}$ power law fits with $\alpha=7.4$. All these curves were obtained for $\tilde{g}N=10^4$ and averaged over the last six oscillation periods.
    }
    \end{center}
\end{figure}

In order to confirm
the robustness of our observations, we 
report here the results of
simulations performed
with a different excitation protocol. Instead of applying a linear oscillating potential, as done in the main text, we consider here
a repulsive Gaussian potential moving from one end of the box to the other:  
$U_\sigma(x,t)=U_0 \exp(-[x-L \sin^2(\Omega t/2)]^2/(2 \sigma^2)) $. This potential can be achieved experimentally using a blue-detuned laser. Once again, we set the oscillation frequency to the value $\Omega = \pi c_0/L$
so that one period of oscillation of the potential corresponds to one round trip at the speed of sound in the box.
Two parameters of this potential can be controlled independently: its height $U_0$, to be compared to the bare chemical potential $\mu_0$, and its width $\sigma$,
chosen as substantially smaller than
the size $L$ of the box. We note that the excitation of the system induced by the potential $U_\sigma$ is stronger when $U_0$ is large and $\sigma$ is small.

By exciting the system with this potential, we observe a behavior which is qualitatively very similar to the one
obtained with the linear oscillating potential of Eq.~\eqref{eq:potential}: large-scale motion with oscillations of the center of mass and the formation
of soliton trains.
  As shown in  Fig.~\ref{fig:other-drive},  the two excitation protocols yield a very similar behaviour of the momentum  distribution and, in both cases, the emergence of the power-law decays discussed in the main text.

\section{Benchmark of soliton counting method}\label{app:counting}

\begin{figure*}[ht]
\begin{center}
   \includegraphics{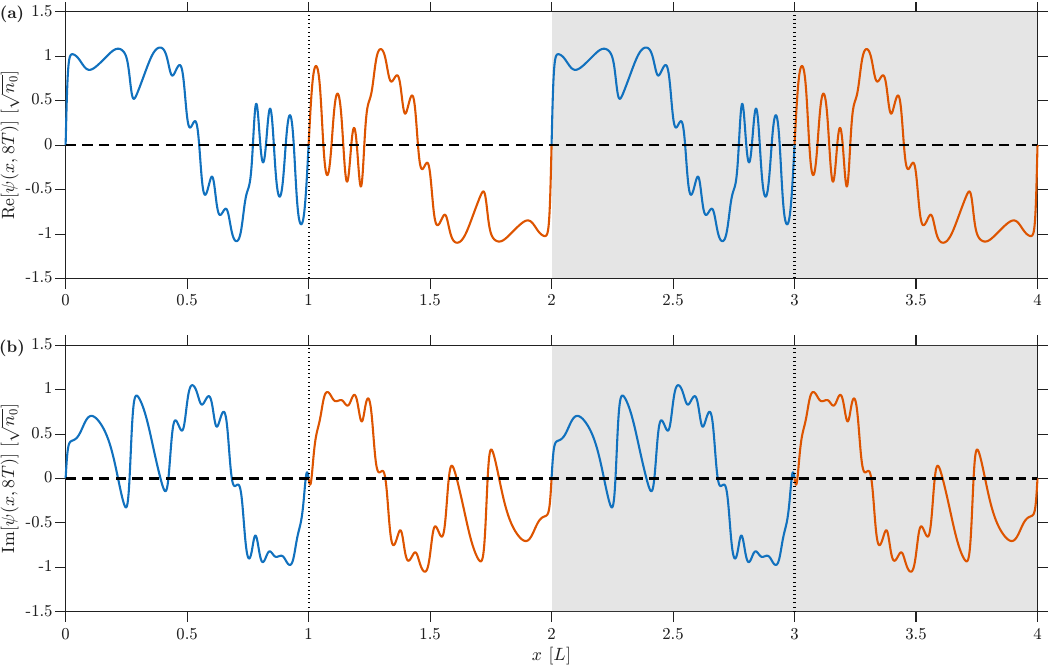}
   \caption{\label{fig:doubling2} (a) Real and (b) imaginary part of the wave function used to identify the eigenvalues of the Lax spectrum corresponding to solitons. The system size is initially $L$. We create an intermediate system of size $2L$ by antisymmetrizing the wave function, hence ensuring periodic boundary conditions. The final system size is $4L$, obtained from two identical copies of the $2L$ system, allowing to double the degeneracy of the eigenvalues of the Lax spectrum corresponding to solitons.}
    \end{center}
\end{figure*}

\begin{figure}[ht]
\begin{center}
    \includegraphics{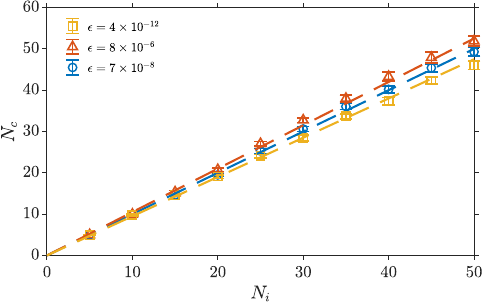}
    \caption{{\label{fig:etalonnage_solitons_diff}} Number of solitons deduced from the counting method $N_c$ as a function of the number $N_i$ of solitons injected in the wave function of the system for different values of the threshold $\epsilon$. 
    The dashed yellow, dashed blue and dashed red lines represent respectively the functions $N_c=0.95N_i$, $N_c=N_i$ and $N_c=1.05N_i$.
    }
    \end{center}
\end{figure}

To identify the eigenvalues of the Lax spectrum corresponding to solitons, we compare the eigenvalues of the intermediate system of size $2L$ ---which allows us to build periodic boundary conditions--- to those of the final system of size $4 L$, see Fig.~\ref{fig:doubling2}.
As explained in the main text, the eigenvalues correspond to solitons when the degeneracy is doubled between the intermediate and final systems. As done in \cite{RomainSciPost}, we build a soliton indicator which takes the value 1 when the eigenvalue of the Lax spectrum corresponds to a soliton and zero if not, of the form:
\begin{equation}
S(\zeta_i)=\frac{\textrm{card}\{\zeta_j' \textrm{ such that }|\zeta_i-\zeta_j'|< \epsilon \}}{\textrm{card}\{\zeta_j \textrm{ such that }|\zeta_i-\zeta_j|< \epsilon \}}-1
\end{equation}
with $\zeta_i$ the set of eigenvalues of the Lax spectrum of the system of size $2L$ and $\zeta_i'$ the set of eigenvalues of the system of size $4L$. The parameter $\epsilon$ corresponds to a threshold between degenerate and non-degenerate eigenvalues to take into account the numerical errors. To benchmark the threshold $\epsilon$, we calculate the Lax spectrum on test wave functions containing a known number of solitons. To build such wave functions, we multiply the wave function at equilibrium by the wave functions of several, evenly spread single solitons given by equation (\ref{eq:fonction_soliton}), with a density $n_s$ low enough to stay in the limit of the dilute soliton regime, and we draw randomly 
the normalized speeds $\nu$ with a uniform law. We applied the soliton counting method for a number of imprinted solitons between 5 and 50, with twenty different test wave functions for each given number of solitons. We find that the threshold which minimizes the relative difference between the number of solitons injected and the number of solitons detected by the soliton indicator is $\epsilon_0 = 7.10^{-8}$. To estimate uncertainties, we also calculated the threshold that gives a relative error on the number of solitons of $+5\%$, $\epsilon_p = 8.10^{-6} $, and the one that gives a relative error on the number of solitons of $-5\%$, $\epsilon_m = 4.10^{-12}$ (see Fig.~\ref{fig:etalonnage_solitons_diff}). We have used these thresholds $\epsilon_{p,m}$ to set the errors bars in Figs.~\ref{fig:debut_k2}, \ref{fig:queue_dis_xieff}, \ref{fig:etalonnage_solitons_diff} and \ref{fig:ls_over_xiLax}.

\section{Estimate for the effective healing length}
\label{app:effective_xi}

\begin{figure}[t]
\centering
    \includegraphics{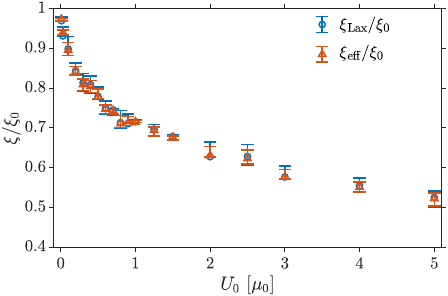}
    \caption{
    Estimate of the effective healing length as a function of the drive strength $U_0/\mu_0$, in units of the healing length $\xi_0$ computed with the initial density $n_0=N/L$. Blue circles: healing length $\xi_{\rm Lax}=\hbar/(\sqrt{2}mc_{\rm Lax})$ obtained from the width of the soliton region in the Lax spectrum. These data are identical to those displayed in Fig.~\ref{fig:queue_dis_xieff} (blue circles). Red triangles: healing length $\xi_{\rm eff}$ as obtained from computing the average volume occupied by a soliton given the actual distribution of the parameters $\nu$ of the solitons, Eq.~\eqref{eq:xi_eff_from_nu} (see text for details). This estimate agrees with $\xi_{\rm Lax}$ within the error bars, indicating that the modification of the speed of sound is indeed due to an excluded volume effect. See Appendix~\ref{app:counting} for the estimate of the uncertainty.
    \label{fig:ls_over_xiLax}
    }
\end{figure}

We estimate the effective healing length $\xi_{\rm eff}$ and the effective speed of sound $c_{\rm eff}$ from the distribution of solitons in the one-dimensional gas with the aid of a simple model. We first estimate the volume occupied by a centered single soliton of normalized velocity $\nu$. The atomic density in the presence of this soliton reads (see Eq.~\eqref{eq:fonction_soliton}):
$$
n(x)=n_\infty\left[(1-\nu^2)\tanh^2\left(\frac{x}{\xi_s}\sqrt{\frac{1-\nu^2}{2}}\right)+\nu^2\right]
$$
where $\xi_s$ characterizes the size of the soliton, that will be eventually related to the density. $n_\infty$ is the atomic density far from the soliton, which is higher than $n_0=N/L$ because of the volume occupied by the soliton. The total atom number is $N=\int_{-L/2}^{L/2} dx \, n(x)$.

By performing  the integral of the density  over the size of the box we obtain:
$$
N= n_\infty \left[L - \xi_s \sqrt{8(1-\nu^2)}\tanh\left(\frac{L}{\xi_s}\sqrt{\frac{1-\nu^2}{8}}\right)\right].
$$
The density $n_\infty$ is thus the one of a uniform distribution in a box of reduced size $L-\Delta L$, with
\begin{equation}
\Delta L = \xi_s \sqrt{8(1-\nu^2)}\tanh\left(\frac{L}{\xi_s}\sqrt{\frac{1-\nu^2}{8}}\right)
\label{eq:soliton_volume}    
\end{equation}
the volume occupied by the soliton. We note that if $L\sqrt{1-\nu^2}\gg \sqrt{8}\xi_s$, which is valid except for the fastest solitons, we have
\begin{equation}
\Delta L\simeq \sqrt{8(1-\nu^2)}\,\xi_s.
\label{eq:soliton_volume_approx}
\end{equation}
For the fastest solitons verifying $L\sqrt{1-\nu^2}\ll \sqrt{8}\xi_s$, we get instead $\Delta L\simeq (1-\nu^2)L$, very small as compared to $\sqrt{8(1-\nu^2)}\,\xi_s$ from our assumption.

In the case of a distribution of $N_s$ solitons, with a given distribution of their parameter $\nu$, the average volume $L_s$ occupied by a soliton in this distribution is obtained by averaging $\Delta L$ over the distribution of $\nu$
\begin{equation}
    L_s = \langle \Delta L \rangle \simeq \sqrt{8}\langle\sqrt{1-\nu^2}\rangle\xi_s.
    \label{eq:soliton_mean_volume_approx}
\end{equation}
The approximation holds if we can neglect the fastest solitons in the distribution, otherwise we should stick to the expression of $\Delta L$ given at Eq.~\eqref{eq:soliton_volume}. In the case of a uniform soliton distribution, we can derive an approximate analytical expression $L_s\simeq \pi \xi_s/\sqrt{2}$, giving an excellent estimate for $L_s$ at large $\tilde{g}N$.

Let us now estimate the effective healing length $\xi_{\rm eff}=\xi_s$ at the background density for our system with $N_s$ solitons in a box, based on Eqs.~\eqref{eq:soliton_volume}, \eqref{eq:soliton_volume_approx} and \eqref{eq:soliton_mean_volume_approx}.
Due to the excluded volume, the effective background density is then approximately
\begin{equation}
  n_{\rm eff} = \frac{N}{L-N_sL_s}=\frac{n_0}{1-n_sL_s},
  \end{equation}
yielding an effective speed of sound
\begin{equation}
    c_{\rm eff} = \frac{c_0}{\sqrt{1-n_s L_s}}.
\end{equation}
$L_s=\lambda\xi_{\rm eff}$ is related to $\xi_{\rm eff}$ through $\lambda\simeq\sqrt{8}\langle\sqrt{1-\nu^2}\rangle$, see Eq.~\eqref{eq:soliton_mean_volume_approx}, with $\lambda\simeq\pi/\sqrt{2}$ if the distribution of the soliton velocities is uniform. Using $\xi_{\rm eff}\propto 1/c_{\rm eff}$, the effective healing length can be deduced self consistently through
\begin{equation}
    \xi_{\rm eff} = \xi_0\sqrt{1-\lambda n_s \xi_{\rm eff}}.
\end{equation}
We find
\begin{equation}
    \xi_{\rm eff} = \xi_0\left[\sqrt{1+\frac{\lambda^2n_s^2\xi_0^2}{4}} - \frac{\lambda n_s\xi_0}{2}\right]
\end{equation}
or equivalently, using $n_s\xi_0=N_s/\sqrt{2\tilde{g}N}$,
\begin{equation}
    \xi_{\rm eff} = \xi_0\left[\sqrt{1+\frac{\lambda^2N_s^2}{8\tilde{g}N}} - \frac{\lambda N_s}{\sqrt{8\tilde{g}N}}\right].
\end{equation}
Using $\lambda\simeq\sqrt{8}\langle\sqrt{1-\nu^2}\rangle$, we finally obtain
\begin{equation}
    \frac{\xi_{\rm eff}}{\xi_0} \simeq \sqrt{1+\frac{N_s^2\langle\sqrt{1-\nu^2}\rangle^2}{\tilde{g}N}} - \frac{ N_s\langle\sqrt{1-\nu^2}\rangle}{\sqrt{\tilde{g}N}}.
    \label{eq:xi_eff_from_nu}
\end{equation}

In Fig.~\ref{fig:ls_over_xiLax}, we compare this expression of $\xi_{\rm eff}$ to $\xi_{\rm Lax}=\hbar/(\sqrt{2}mc_{\rm Lax})$ obtained from the speed of sound $c_{\rm Lax}$ deduced from the width of the discrete region in the Lax spectrum. We find an excellent agreement, supporting our interpretation of the increase of the speed of sound and the decrease of the healing length from an excluded-volume effect.

\end{document}